      [1999/12/01 v1.4c Il Nuovo Cimento]
\documentclass{cimento}

\usepackage{graphicx}  % got figures? uncomment this
\usepackage{epstopdf}
\usepackage{xspace}
\title{LHCb status and early physics prospects}
\author{Monica~Pepe~Altarelli\from{ins:x}, on behalf of the $\rm{LHCb}$ Collaboration}
\instlist{\inst{ins:x} CERN, PH Department, CH-1211 Geneve 23, Switzerland\\
Monica.Pepe.Altarelli@CERN.CH}
\PACSes{
\PACSit{12.15.Ff, 12.15.Hh, 12.60, 13.20.He, 14.65Fy}{\ldots}}
\begin{document}

%%%%%   macros for  LHCb 
%%
%% Luminosity
\newcommand{\lum}{\mathcal{L}}
\newcommand{\lumin}[2]{{#1}\;10^{#2}\;\mathrm{cm^{-2} s^{-1}}}
\newcommand{\lolumi}{\mbox{$2 \times 10^{32}$~cm$^{-2}$ s$^{-1}$~}}
\newcommand{\hilumi}{\mbox{$5 \times 10^{32}$~cm$^{-2}$ s$^{-1}$~}}
%% Units 
\newcommand{\cms}{\mbox{cm$^{-2}$ s$^{-1}$~}}
\newcommand{\mq}{\ensuremath{\; \mathrm{m^2}}}
\newcommand{\cmq}{\ensuremath{\; \mathrm{cm^2}}}
\newcommand{\ccm}{\ensuremath{\; \mathrm{cm^3}}}
\newcommand{\mmq}{\ensuremath{\; \mathrm{mm^2}}}
\newcommand{\us}{\; \mu\:\mathrm{s}}
\newcommand{\gra}{{\mathrm{^o}}}
\newcommand{\bb}{$\rm b\overline{b}$}
\newcommand{\invcmq}{\ensuremath{\mathrm{/{cm^2}}}}
\newcommand{\mHzpercmq}{\ensuremath{\mathrm{MHz/cm^2}}}
\newcommand{\kHzpercmq}{\ensuremath{\mathrm{kHz/cm^2}}}
\newcommand{\Hzpercmq}{\ensuremath{\mathrm{Hz/cm^2}}}
\newcommand{\Cpercm}{\ensuremath{\mathrm{C/cm}}}
\newcommand{\Cpercmq}{\ensuremath{\mathrm{C/cm^2}}}
\newcommand{\ItwoC}{\ensuremath{\mathrm{I^2C}}}
\newcommand{\x}{\ensuremath{\times}}
\newcommand{\microns}{\ensuremath{\mathrm{\mu m}}}

\def\GeVc{GeV$/c$\xspace}
\def\GeVcc{GeV$/c^2$\xspace}
\def\MeVcc{MeV$/c^2$\xspace}
\def\PT{p_{\mathrm{T}}\xspace}
\def\ET{E_{\mathrm{T}}\xspace}
%%%
\def\Y#1S{{\Upsilon\mathrm{(#1S)}}}
\def\pipi{\pi^+\pi^-}
\def\ubar{\bar u}
\def\dbar{\bar d}
\def\sbar{\bar s}
\def\cbar{\bar c}
\def\bbar{\bar b}
\def\tbar{\bar t}
\def\Kbar{\rm {\kern 0.2em\overline{\kern -0.2em K}}{}}
\def\Bbar{\rm {\kern 0.18em\overline{\kern -0.18em B}}{}}
\def\Dbar{\rm {\kern 0.2em\overline{\kern -0.2em D}}{}}
\def\pbar{\rm {\kern 0.06em\overline{\kern -0.06em p}}{}}
\def\Lambdabar{\rm {\kern 0.06em\overline{\kern -0.06em \Lambda}}{}}
\def\Jpsi{\ensuremath{\rm J\mskip -3mu/\mskip -2mu\psi\mskip 2mu}}
\def\Ks{\rm {K^0_S}}
\def\Kl{\rm {K^0_L}}
\def\BS{$\rm B_s$\xspace}
\def\BC{$\rm B_c$\xspace}
\def\BBar{$\rm {B^0_s}$-${\rm {\kern 0.18em\overline{\kern -0.18em B^0_s}}{}}$\xspace}
%%%%%%
\def\BJMUK{$\rm B^0_d \rightarrow \Jpsi(\mu^+ \mu^-) K^0_S$}
\def\BJMUKVIS{$\rm B^0_d \rightarrow \Jpsi(\mu^+ \mu^-) K^0_S(\pi^+ \pi^-)$}
\def\BKMUMU{$\rm B_d \rightarrow K^{*0}\mu^+ \mu^-$}
\def\BJMUKBOLD{$\bfmath{\rm B^0_d \rightarrow \Jpsi(\mu^+ \mu^-) K^0_S}$}
\def\BJPSIPHI{$\rm B^0_s \rightarrow \Jpsi \phi$\xspace}
\def\BJPHI{$\rm B^0_s \rightarrow \Jpsi \phi$\xspace}
\def\BSMUMU{$\rm B^0_s \rightarrow \mu^+ \mu^-$\xspace}
\def\BSPHIPHI{$\rm B^0_s \rightarrow \phi\phi$\xspace}
\def\BSMUMUBOLD{$\bfmath{\rm B^0_s \rightarrow \mu^+ \mu^-}\xspace$}
\def\BJPSIKS{$\rm B^0_d \rightarrow \Jpsi Ks $}
\def\BMUX{$\rm b \rightarrow \mu\ X$}
\def\pkm{$\rm \pi,K \rightarrow \mu\ \nu$}
%%
%Gas
\def\isob{$\rm C_4H_{10}$\xspace}
\def\SF6{$\rm SF_6$\xspace}
\def\CF4{$\rm CF_4$\xspace}

\let\bra\langle

\let\ket\rangle

\maketitle

\begin{abstract}
LHCb is a dedicated detector for  b and c physics at the LHC. I will present a concise review of the detector design and expected performance together with some first results on the commissioning of the different sub-systems based on cosmic data and particle beams delivered by the LHC during the summer of 2008. The experiment is ready to exploit first data expected from the LHC. An integrated luminosity of $\sim 0.3~\rm{fb}^{-1}$, which should be collected during the first year of physics running, will already allow LHCb to perform a number of very significant mesurements with the potential of revealing New Physics effects, such as the measurement of the \BS mixing phase $\phi_{\Jpsi\phi}$, or the search for the decay \BSMUMU beyond the limit set by CDF and D0.
\end{abstract}

\section{Introduction}
LHCb is a dedicated b and c-physics precision experiment at the LHC that will search for New Physics (NP) beyond the Standard Model (SM) through the study of very rare decays of charm and beauty-flavoured hadrons and precision measurements of CP-violating observables.  In the last decade, experiments at B factories have confirmed that the mechanism proposed by Kobayashi and Maskawa is the major source of CP violation  observed so far. The SM description of flavour-changing processes has been confirmed in the b$\rightarrow$d transition at the level of 10-20\% accuracy. However, NP effects can still be large in b$\rightarrow$s transitions, modifying the \BS mixing phase   $\phi_{\Jpsi\phi}$, measured from \BJPSIPHI decays, or in channels dominated by other loop diagrams, such as the very rare decay \BSMUMU, or in \BSPHIPHI. Therefore, the challenge of the future b experiments is to widen the range of measured decays, reaching channels that are strongly suppressed in the SM and, more generally, to improve the precision of the measurements to achieve the necessary sensitivity to NP effects in loops. LHCb will extend the b-physics results from the B factories by studying decays of heavier b hadrons, such as \BS or \BC, which will be copiously produced at the LHC. It will complement the direct search of NP at the LHC by providing important information on the NP flavour structure through a dedicated detector, optimized for this kind of physics.

\section{b physics at the LHC: environment, background, general trigger issues }
The LHC will be the world's most intense source of b hadrons. In proton-proton collisions at  $\sqrt{s}=14$~TeV, the  $\rm {b\bbar}$ cross section is expected to be $\sim 500~ \mu$b producing $10^{12}$ $\rm {b\bbar}$ pairs in a standard ($10^7$s) year of running at the LHCb operational luminosity of  \lolumi. As in the case of the Tevatron, a complete spectrum of b hadrons will be available, including \BS, \BC mesons and baryons such as $\Lambda_{\rm b}$. However, less than 1\% of all inelastic events contain b quarks, hence triggering is a critical issue.

At the nominal LHC design luminosity of $10^{34}$\cms, multiple p-p collisions within the same bunch crossing (so-called pile-up) would significantly complicate the b-production and decay-vertex reconstruction. For this reason the luminosity at LHCb will be locally controlled by appropriately focusing the beam to yield  $\lum = 2-5\times10^{32}\cms$, at which the majority of the events have a single p-p interaction. This matches well with the expected LHC conditions during the start-up phase. Furthermore, running at relatively low luminosity reduces the detector occupancy of the tracking systems and limits radiation damage effects. 

The dominant  production mechanism at the LHC is through gluon-gluon fusion in which the momenta of the incoming partons are strongly asymmetric in the p-p centre-of-mass frame. As a consequence, the $\rm {\bbar}$ pair is boosted along the direction of the higher momentum gluon, and both b hadrons are produced in the same forward (or backward) direction in the p-p centre-of-mass frame. The detector is therefore designed as a single arm forward spectrometer covering the pseudorapidity range $1.9<\eta<4.9$ , which ensures a high geometric efficiency for detecting all the decay particles from one b hadron together with the decay particles from the accompanying $\rm {\bbar}$ hadron to be used as a flavour tag. A modification to the LHC optics, displacing the interaction point by 11.25 m from the centre, has permitted maximum use to be made of the existing cavern by freeing 19.7 m for the LHCb detector components. 

A detector design based on a forward spectrometer offers further advantages: b hadrons are expected to have a hard momentum spectrum in the forward region; their average momentum is  $\sim$80~\GeVc, corresponding to approximately 7~mm mean decay distance, which facilitates the separation between primary and decay vertices. This property, coupled to the excellent vertex resolution capabilities, allows proper time to be measured with a resolution of $\sim$40 fs, which is crucial for studying CP violation and oscillations with \BS mesons, because of their high oscillation frequency. Furthermore, the forward, open geometry allows the vertex detector to be positioned very close to the beams and facilitates detector installation and maintenance. In particular, the silicon detector sensors, housed, like Roman pots, in a secondary vacuum, are split in two halves that are retracted by   $\sim$30 mm from the interaction region before the LHC ring is filled, in order to allow for beam excursions during injection and ramping. They are then positioned within $\sim$8 mm from the interaction region after stable beam conditions have been obtained. 

\begin{figure} [hbt]
\centering
\includegraphics [width=6.0 cm]{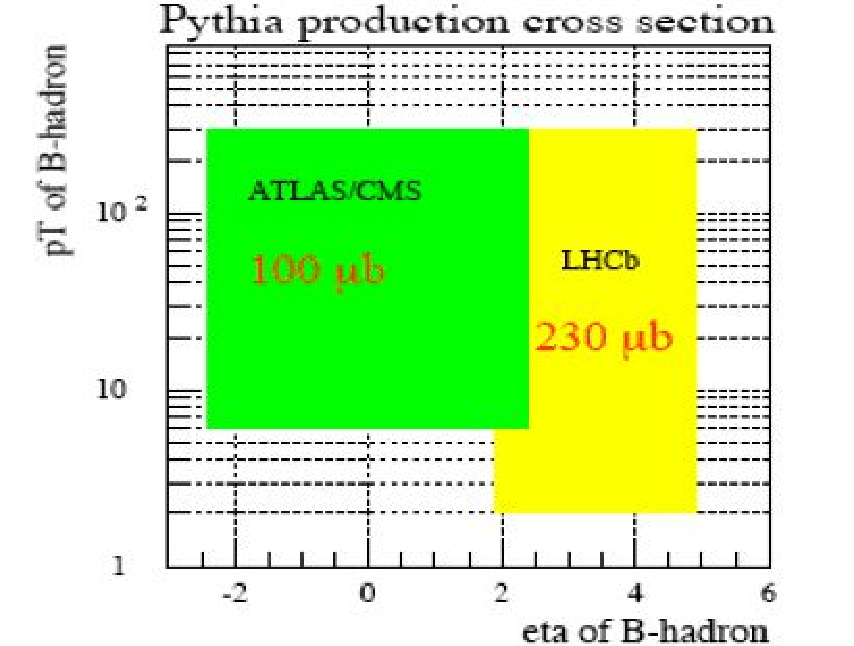}    
\caption{b-hadron transverse momentum  as a function of the pseudorapidity  $\eta$, showing the  $(\eta,\PT)$ regions covered by ATLAS and CMS, compared to that covered by LHCb.\label {eta}}
\end{figure}

Figure~\ref{eta} illustrates the LHCb acceptance in the plane $(\eta,\PT)$ of the b hadrons in comparison to that of ATLAS and CMS:  ATLAS and CMS cover a pseudorapidity range of  $|\eta|<2.5$ and rely on high-$\PT$ lepton triggers. LHCb relies on much lower $\PT$ triggers, which are efficient also for purely hadronic decays. Most of the ATLAS and CMS b-physics programme will be pursued during the first few years of operation, for luminosities of order $10^{33}$\cms. Once LHC reaches its design luminosity, b physics will become exceedingly difficult for ATLAS and CMS due to the large pile-up (~20 interactions per bunch crossing, on average), except for very few specific channels characterized by a simple signature, like \BSMUMU. 
\section{Detector description and performance}
\begin{figure} [htb]
\centering
\includegraphics [width=10.0 cm]{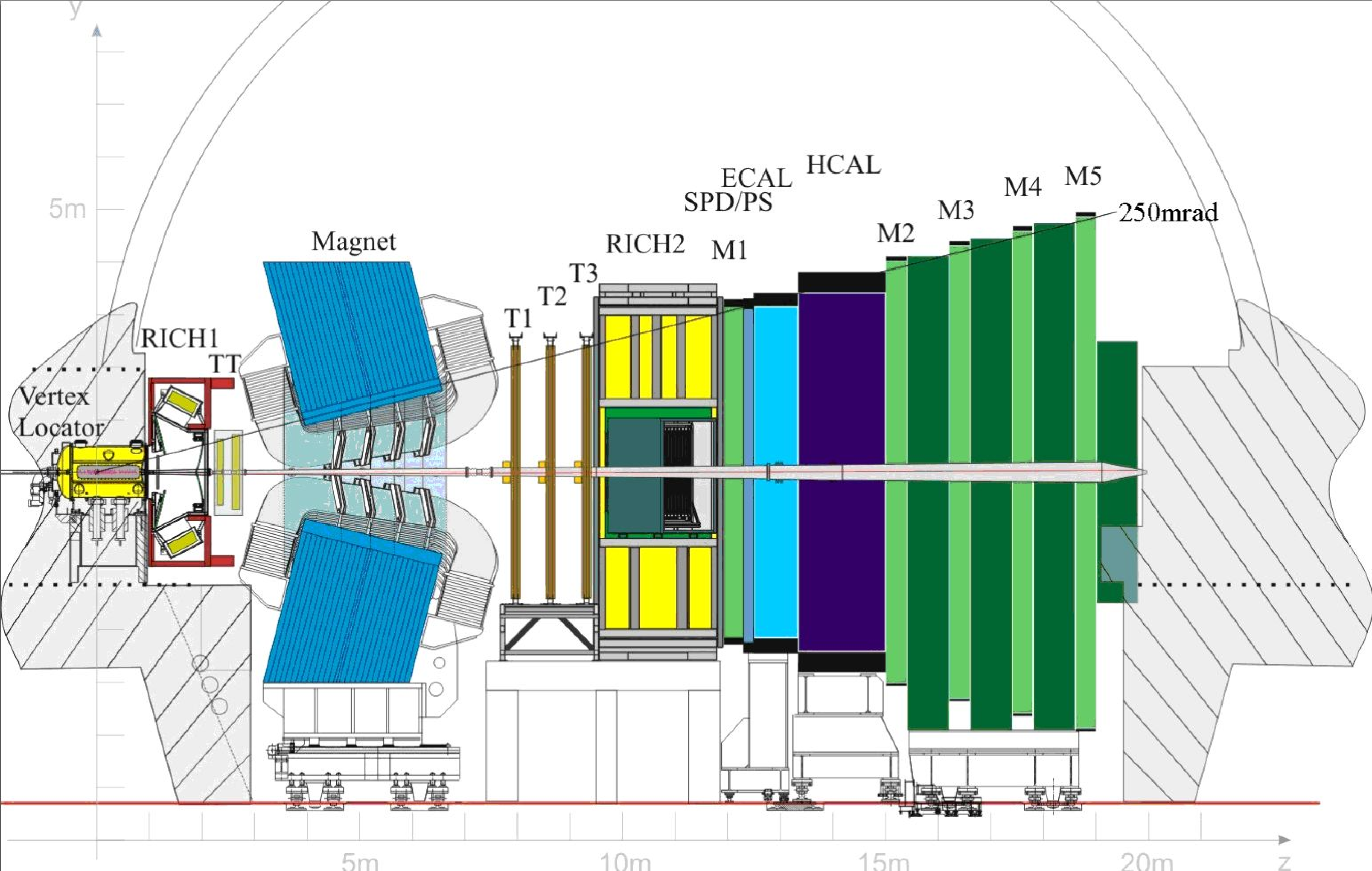}    
\caption{Side view of the LHCb detector showing the Vertex Locator (VELO), the dipole magnet, the two RICH detectors, the four tracking stations TT and T1-T3, the Scintillating Pad Detector (SPD), Preshower (PS), Electromagnetic (ECAL) and Hadronic (HCAL) calorimeters, and the five muon stations M1-M5.\label{detector}}
\end{figure}

The key features of the LHCb detector are: 
\begin{itemize}
\item A versatile trigger scheme efficient for both leptonic and hadronic final states, which is able to cope with a variety of modes with small branching fractions;
\item Excellent vertex and proper time resolution;
\item Precise particle identification (ID), specifically for hadron ($\pi$/K) separation;
\item Precise invariant mass reconstruction to efficiently reject background due to random combinations of tracks. This implies a high momentum resolution.
\end{itemize}
A schematic layout is shown in fig.~\ref{detector}. It consists of a vertex locator (VELO), a charge particle tracking system with a large aperture dipole magnet, aerogel and gas Ring Imaging Cherenkov counters (RICH), electromagnetic (ECAL) and hadronic (HCAL) calorimeters and a muon system.
In the following, the most salient features of the LHCb detector are described in more detail. A much more complete description of the detector characteristics can be found in \cite{ref:reopt, ref:det}.
\subsection {Trigger}
One of the most critical elements of LHCb is the trigger system. At the chosen LHCb nominal luminosity, taking into account the LHC bunch crossing structure, the rate of events with at least two particles in the LHCb acceptance is $\sim$10~MHz (instead of the nominal 40 MHz LHC crossing rate). The rate of events containing b quarks is $\sim$100~kHz, while the rate of events containing c quarks is much larger ($\sim$ 600~kHz). However, the rate of ÔinterestingÕ events is just a very small fraction of the total rate ($\sim$Hz), due to the combined effect of branching fraction and detector acceptance, hence the need for a highly selective and efficient trigger.

The LHCb trigger exploits the fact that b hadrons are long-lived, resulting in well separated primary and secondary vertices, and have a relatively large mass, resulting in decay products with large $\PT$. It consists of two levels: Level0 (L0) and High Level Trigger (HLT). L0, implemented on custom electronics boards, is designed to reduce the input rate to 1 MHz at a fixed latency of 4~$\mu$s. At this rate, events are sent to a computer farm with up to $\sim$2000 multi-processor boxes where several HLT software algorithms are executed. The HLT, which has access to the full detector information, reduces the rate from 1 MHz to $\sim$2~kHz.

L0, based on calorimeter and muon chamber information, selects muons, electrons, photons or hadrons above a given $\PT$ or $\ET$ threshold, typically in the range 1 to 4 GeV.  The L0 hadron trigger occupies most of the bandwidth (~700 kHz) and is unique within the LHC experiments. The muon triggers (single and double) select $\sim$200 kHz, while the rest of the bandwidth is due to the electromagnetic calorimeter triggers. Typically, the L0 efficiency is $\sim$50\% for hadronic channels, $\sim$90\% for muon channels and $\sim$70\% for radiative channels, normalized to offline selected events.

The HLT algorithms are designed to be simple, to minimize systematic uncertainties, and fast. This is realized by reconstructing for each trigger only a few tracks, which are used for the final decision. The HLT comprises several paths (alleys) to confirm and progressively refine the L0 decision, followed by inclusive and exclusive selections. The choice of the alley depends on the L0 decision. The average execution time is few ms, which matches with the expected size of the CPU farm. The total trigger rate after the HLT is $\sim$2 kHz, a relatively high rate that also includes calibration samples to be used to understand the detector performance. 
\subsection{ VELO and Tracking System}		
The LHCb tracking system consists of a warm dipole magnet, which generates a magnetic field integral of $\sim$4~Tm, four tracking stations and the VELO. The first tracking station located upstream of the magnet consists of four layers of silicon strip detectors. The remaining three stations downstream of the magnet are each constructed from four double-layers of straw tubes in the outer region, covering most ($\sim$98\%) of the tracker area, and silicon strips in the area closer to the beam pipe ($\sim$2\%). However, $\sim$20\% of the charged particles traversing the detector go through the silicon inner tracker, due to the forward-peaked multiplicity distribution. The expected momentum resolution increases from $\delta\rm{p/p}\sim0.35\%$ for low momentum tracks to 0.55\% at the upper end of the momentum spectrum. This translates into an invariant mass resolution of $\delta\rm{M}\sim20~MeV/c^2$ for $\rm{B_s}$ decays into two charged tracks, such as \BSMUMU, substantially better than in the General-Purpose detectors at LHC.

The VELO consists of 21 stations, each made of two silicon half disks, which measure the radial and azimuthal coordinates. The VELO has the unique feature of being located at a very close distance from the beam line (0.8 cm), inside a vacuum vessel, separated from the beam vacuum by a thin aluminum foil. This allows an impressive vertex resolution to be achieved, translating, for instance, in a proper time resolution of $\sim$36~fs for the decay \BJPSIPHI, $\it{i.e.}$ a factor of ten smaller than the \BS oscillation period and a factor of two better than in the General-Purpose detectors. The resolution on the impact parameter can be parameterized as $\delta{IP}\sim14\mu\rm{m} + 35\mu\rm{m}/\PT$.
\subsection{Particle Identification}
Particle identification is provided by the two RICH detectors and the Calorimeter and Muon systems. 

The RICH system is one of the crucial components of the LHCb detector.  The first RICH, located upstream of the magnet, employs two radiators, \isob  gas and aerogel, ensuring a good   separation in the momentum range from 2 to 60~\GeVc. A second RICH in front of the calorimeters, uses a \CF4 gas radiator and extends the momentum coverage up to $\sim$100~\GeVc. The calorimeter system comprises a pre-shower detector consisting of 2.5 radiation length lead sheet sandwiched between two scintillator plates, a 25 radiation length lead-scintillator electromagnetic calorimeter of the shashlik type and a 5.6 interaction length iron-scintillator hadron calorimeter. The muon detector consists of five muon stations equipped with multiwire proportional chambers, with the exception of the centre of the first station, which uses triple-GEM detectors. 

Electrons, photons and $\pi^0$s are identified using the Calorimeter system. The average electron identification efficiency  extracted from $\Jpsi\rightarrow \rm{ e^+e^-}$ decays is $\sim95\%$ for a pion misidentification rate of $\sim0.7\%$. Muons are identified using the muon detector with an average efficiency in the acceptance extracted from $\Jpsi\rightarrow \mu\mu$ decays of $\sim93\%$ for a pion misidentification rate of $\sim1\%$. 
The RICH system provides good particle identification over the entire momentum range. The average efficiency for kaon identification for momenta between 2 and 100~\GeVc is $\epsilon (\rm{K} \rightarrow \rm{K})\sim$95\%, with a corresponding average pion misidentification rate $\epsilon (\pi \rightarrow \rm{K})\sim$5\%.

\section{Commissioning}
\begin{figure} [htb]
\centering
\includegraphics [width=14.0 cm]{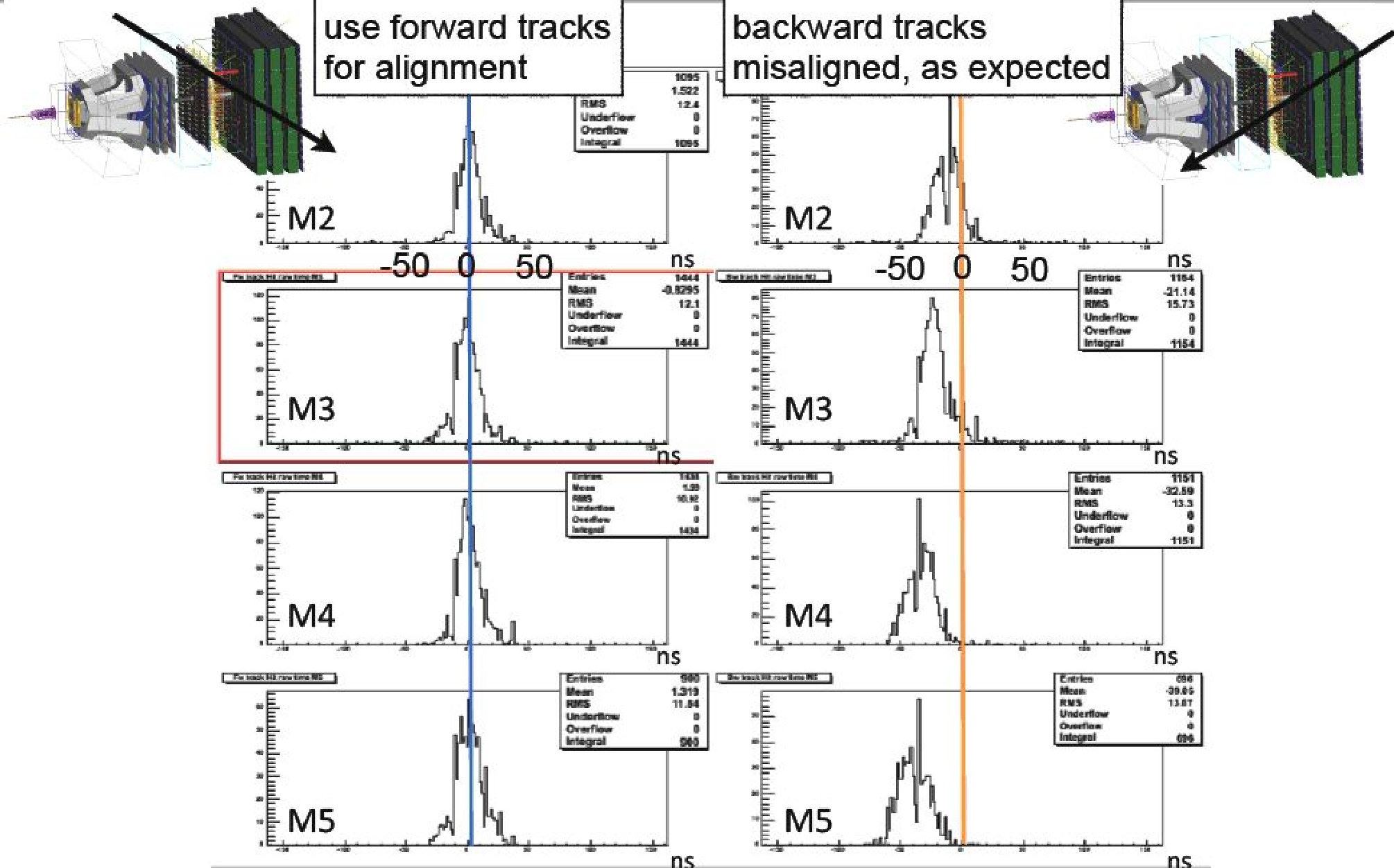}    
\caption{ Readout time distribution with respect to reference time (in ns) for forward and backward tracks in four  muon stations  (M2-M5).\label{cosmic}}
\end{figure}

During the summer 2008 the commissioning activities converged into a fully operational detector. All sub-detectors were included  under central control and data taking was extensively exercised. Although the geometry of the LHCb detector is not well suited for cosmic runs (the rate of 'horizontal' cosmic events is well below 1~Hz), over one million cosmic events were recorded using Muon and Calorimeter cosmic triggers. These cosmic data have proved to be extremely useful to perform a coarse initial geometrical and time alignment of the larger detectors. As an example, fig.~\ref{cosmic} shows  the distributions of the readout time, optimized for forward tracks, of four muon stations. The readout time is measured with respect to the trigger time as provided by the the scintillator pad detector of the calorimeter system. The distributions are nicely centred at zero only for the forward tracks.

At the end of August 2008, the machine carried out several tests of the transfer line ('synchronization tests'). A beam of 450~GeV protons extracted from the SPS was injected into the LHC and dumped on an injection line beam stopper ('TED') located  approximately 340~m downstream of LHCb. This produced a large flux of nearly parallel particles that hit the detector from the back, $\it{i.e.}$ from the muon stations towards the VELO. These data were particularly useful for the time and position alignment of those sub-detectors (IT, TT and VELO) that could not make use of cosmic events, as they are either too small or too distant from the detectors providing the trigger. The initial alignment with reconstructed tracks indicated no major problems and a resolution in the expected range.

As an example, for the VELO the detector displacement was measured to be less than $\sim10~\microns$ with respect to the survey information. The achieved resolution was measured to match the binary resolution ($=\frac{\rm{pitch}}{\sqrt{12}}$), which is consistent with expectation, given that the large majority ($\sim90\%$) of the used clusters are one-strip clusters. Improvements to the resolution are expected with an optimized tuning of the signal processing algorithms and of the readout  time alignment. Moreover, the resolution is expected to be better for tracks at angles around 140 mrad, for which the charge sharing between adjacent strips is optimal.

\section{Early physics at LHCb}
The objective of the very first running phase is to complete the commissioning of the sub-detectors and of the trigger. Large minimum bias data samples ($\sim 10^8$ per day at 2~kHz output rate) will be collected as soon as the LHC delivers p-p collisions using a simple interaction trigger based on total energy in the calorimeters. These data will provide a high-statistics and high-purity $\rm{V}^0$ sample ($\Ks$, $\Lambda$, $\Lambdabar$), which can be used to probe the hadronization process in a rapidity range complementary to that of the other LHC detectors and in which phenomenological models tuned to Tevatron data show significant differences when extrapolated to LHC energies. Measurements will include differential cross sections and production ratios for different strange particles as a function of rapidity and transverse momentum. 

A simple, lifetime-unbiased  muon trigger, requiring $\PT>1$\GeVc at L0, will allow us to collect a large, clean sample of $\Jpsi\rightarrow\mu\mu$ decays. With an integrated luminosity of $5~\rm{pb}^{-1}$,  $\sim3\times10^6$ events are expected  with a S/B of $\sim$4 at $\sqrt{s}=8$~TeV, which may be the initial LHC centre-of-mass energy.  This sample can be used to extract  both the prompt $\Jpsi$ and $\rm{b}\rightarrow\Jpsi$ production cross sections in a region not accessible to other collider experiments. Other charmonia related measurements will also be performed, such as that of the $\Jpsi$ polarization or of the production of the exotic X, Y and Z charmonia states observed in recent years.

An integrated luminosity of $0.2-0.3~\rm{fb}^{-1}$, which should hopefully be collected during the first year of physics running, will already allow LHCb to realize a number of very significant b-physics measurements, with the potential of revealing NP effects, such as the measurement of the \BS mixing phase $\phi_{\Jpsi\phi}$, the search of the decay \BSMUMU beyond the limit set by CDF and D0, or the study of the decay \BKMUMU (for this last decay, with $0.2~\rm{fb}^{-1}$ LHCb should be able to collect $\sim$700 events, which is a larger sample than at all existing facilities combined). 

\begin{figure} [htb]
\centering
\includegraphics [width=7.0 cm]{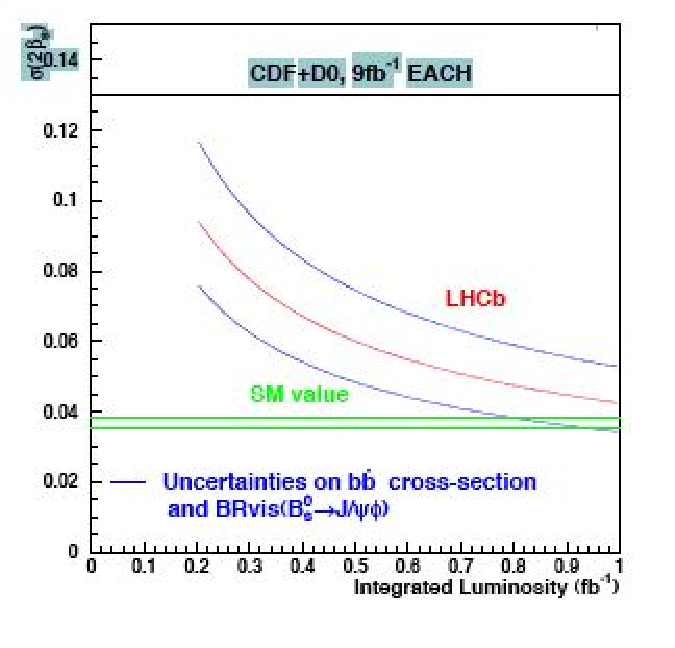}    
\caption{Statistical uncertainty on $\phi_{\Jpsi\phi}$ versus the integrated luminosity. The blue lines show the uncertainties coming from the b$\rm {\bbar}$ cross section and the visible branching ratio on \BJPSIPHI. The green band is the SM value. The Tevatron line is the combined CDF/D0 uncertainty in 2008 scaled to $18~\rm{fb}^{-1}$, as expected  by the end of run 2.\label{betas}}
\end{figure}

A flavour tagged, angular analysis of the decay \BJPSIPHI allows the determination of a CP-violating phase $\phi_{\Jpsi\phi}$. In the SM this phase  is predicted to be  $-2\beta_{\rm s}\simeq-0.04$, where $\beta_{\rm s}$ is the smaller angle of the "b-s unitarity triangle". However NP could significantly modify this prediction, if new particles contribute to the \BBar box diagram. In fact, the CDF and D0 collaborations \cite{ref:CDFphis, ref:D0phis} have reported a first measurement of the \BS mixing phase. Their combined result deviates from the SM prediction by $\sim2.2\sigma$ with a central value for $2\beta_{\rm s}$ as large as 0.77. LHCb has the capability to significantly improve the existing experimental knowledge of this phase thanks to the large signal yield ($\sim$12k events for $0.2~\rm{fb}^{-1}$), the excellent proper time resolution to resolve fast \BS oscillations ($\sim$40~fs), the good flavour tagging ($\sim6\%$), and the good control of the proper time and angular acceptance. Figure \ref{betas} shows the statistical uncertainty on the phase $\phi_{\Jpsi\phi}$ as a function of the integrated luminosity. As one can see, already with $0.3~\rm{fb}^{-1}$,  LHCb should  be able to improve on the expected Tevatron limit.

\begin{figure} [tbh]
\centering
\includegraphics [width=7.0 cm]{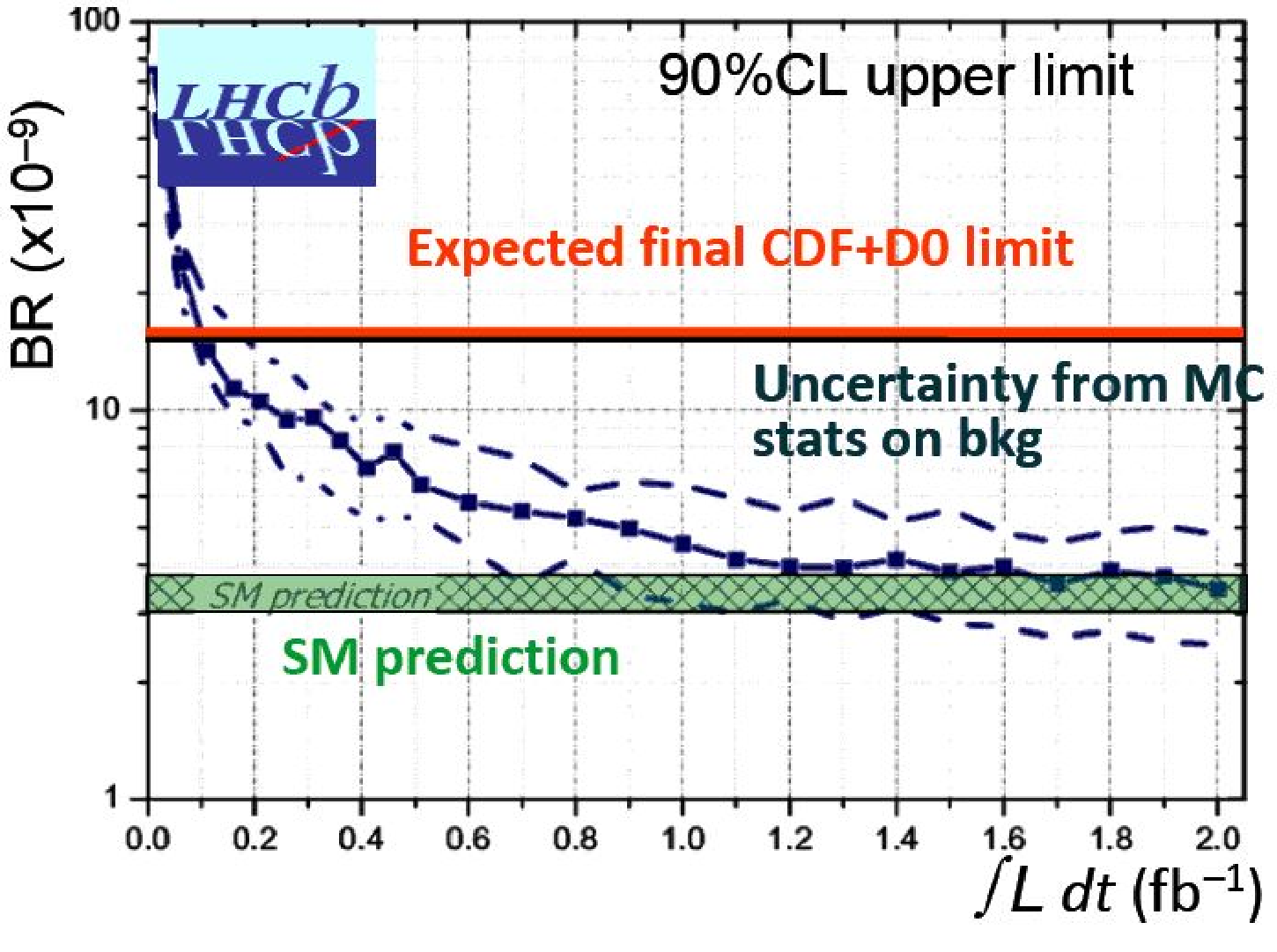}    
\caption{Expected 90\% CL upper limit of BR(\BSMUMU) in absence of signal as a function of the integrated luminosity. The green band is the SM value. The Tevatron line is the combined CDF/D0 uncertainty in 2008 scaled to $18~\rm{fb}^{-1}$, as expected by the end of run 2.\label{bsmumu}}
\end{figure}

The decay \BSMUMU has been identified as another very interesting potential constraint on the parameter space of models for physics beyond the SM. The BR for this decay is computed to be very small in the SM: $\rm{BR}$(\BSMUMU)$=(3.35\pm0.32)\times{10^{-9}}$\cite{ref:bsmumuth}, but could be enhanced in certain NP scenarios. For example, in the MSSM, this branching ratio is known to increase as the sixth power of $\rm{tan}\beta=\nu_{\rm u}/\nu_{\rm d}$, the ratio of the two vacuum expectation values. Any improvement to this limit is therefore particularly important for models with large $\rm{tan}\beta$. The upper limit to the \BSMUMU branching ratio measured at the Tevatron is $\rm{BR}$(\BSMUMU)$<4.7\times 10^{-8}$ at 90\%CL\cite{ref:CDFbsmumu, ref:D0bsmumu}. For this measurement,  LHCb has developed an analysis based on the use of control channels to minimize the dependence on MC simulation.  The main issue for this analysis is the rejection of the background, largely dominated  by random combinations of two muons originating from two distinct b decays. This background can be kept under control by exploiting the excellent LHCb vertexing capabilities, and mass resolution. Figure \ref{bsmumu} shows, as a function of the integrated luminosity, the BR value excluded at 90\%CL. Already with $\sim0.2~\rm{fb}^{-1}$ LHCb should improve on the expected Tevatron limit, while  a $\sim3\sigma$ observation will require an integrated luminosity of $3~\rm{fb}^{-1}$, assuming the SM value.

In subsequent years, the experiment will develop its full physics programme, and plans are to accumulate an integrated luminosity of $\sim10~\rm{fb^{-1}}$. Such a data sample will, for example, allow LHCb to improve the error on the CKM angle $\gamma$ by a factor of $\sim$five, and probe NP in rare B meson decays with electroweak, radiative and hadronic penguin modes.
\section{Conclusions}
The large $\rm {b\bbar}$ production cross section at the LHC provides a unique opportunity to study in detail CP violation and rare b decays with the LHCb detector. In particular, production of \BS mesons could play a crucial role in disentangling effects originating from NP and a few observables sensitive to NP should already be accessible at the end of the 1st year of data taking. During the last year LHCb was fully installed and commissioned using cosmic events and first LHC-induced tracks. At this point  we are all eagerly waiting to exploit first data expected from the LHC in 2009/2010.  
\acknowledgments
I would like to thank the Organizers of Les Rencontres for their kind invitation and my LHCb colleagues for providing the material discussed in this article, in particular Tatsuya Nakada for taking the time to read this manuscript.


\begin{thebibliography}{0}
\bibitem{ref:reopt}\BY{$\rm{LHCb}$ Collaboration.} \TITLE{LHCb Reoptimized Detector, Design and Performance},  LHCC-2003-030, 2003;
\bibitem{ref:det}\BY{$\rm{LHCb}$ Collaboration.} \TITLE{The LHCb Detector at the LHC}, Journal of Instrumentation, 2008 JINST 3 S08005;
\bibitem{ref:CDFphis}\BY{CDF Collaboration.} \TITLE{First Flavor-Tagged determination of Bounds on Mixing-Induced CP Violation in \BJPSIPHI Decays},\IN{Phys. Rev. Lett.}{100}{2008}{161802}; 
\bibitem{ref:D0phis}\BY{D0 Collaboration.} \TITLE{Measurement of \BS mixing parameters from the flavor-tagged decay \BJPSIPHI}, arXiv:0802.2255v1 [hep-ex] 2008;
\bibitem{ref:bsmumuth}\BY{Blanke~M., Buras.~A., Guadagnoli D. \atque Tarantino~C.} arXiv:hep-ph/0604057v5 2006;
\bibitem{ref:CDFbsmumu}\BY{CDF Collaboration.} \IN{Phys. Rev. A}{100}{2008}{101802}; 
\bibitem{ref:D0bsmumu}\BY{D0 Collaboration.} D0 note 5344-CONF, 2007.

\end{thebibliography}
\end{document}